\documentclass[showpacs,aps,floatfix,prb,reprint,superscriptaddress]{revtex4-1}
\usepackage{graphicx}
\usepackage{xfrac}
\usepackage{amssymb}
\usepackage{tikz}
\usepackage{amsfonts} 
\usepackage{amsmath} 
\usepackage[utf8]{inputenc} 
\usepackage{mathtools}
\usepackage{xcolor,colortbl}
\usepackage{bm} 
\usepackage{array}
\usepackage{color}  
\usepackage{relsize} 
 \usepackage[english]{babel}
 \usepackage{amssymb,amsmath}
 \usepackage{color}
 \usepackage{pstricks}
 \usepackage{bbold}
 \usepackage{graphicx}
 \usepackage{color}
 \usepackage{verbatim}
 \usepackage{fancyvrb}
 \usepackage{alltt}
 \usepackage{longtable}
 \usepackage{amsmath}

\begin{document}

\title{Plasmonic properties of refractory titanium nitride}

\author{Alessandra Catellani}
\affiliation{CNR-NANO Research Center S3, Via Campi 213/a, 41125 Modena, Italy}

\author{Arrigo Calzolari}
\email {Email: arrigo.calzolari@nano.cnr.it}
\affiliation{CNR-NANO Research Center S3, Via Campi 213/a, 41125 Modena, Italy}

\date{\today}

\begin{abstract}
The development of plasmonic and metamaterial devices requires the research of 
high-performance materials, alternative to standard noble metals. 
Renewed as refractory stable compound for durable coatings, titanium nitride has been recently proposed as an efficient plasmonic  material.
Here, by using a first principles approach,  we investigate the  
plasmon dispersion relations of TiN bulk 
and  we predict the effect of pressure on its optoelectronic properties.
Our results 
explain the main features of TiN in the visible range and prove a universal scaling law
which relates its mechanical and plasmonic properties as a function of pressure.  
Finally, we  address the formation and stability of  surface-plasmon polaritons at 
different TiN/dielectric interfaces proposed by recent experiments.
The unusual combination of plasmonics and refractory features paves the way for the realization of plasmonic devices able to work at conditions not sustainable by usual noble metals.
\end{abstract}

\maketitle

\section{Introduction} 
Plasmons, i.e. collective excitations of the electron density, control the high frequency electromagnetic response of materials
as they affect their intrinsic dielectric and refractive properties, along with the screening and the amplification of the external fields at  surface boundaries.\cite{Maier:2007wq, murray07}
Plasmonic resonances can couple with electromagnetic fields and either propagate along extended interfaces as Surface Plasmon Polaritons (SPPs) or being localized 
at nanoscale surfaces as Localized Surface Plasmon Resonances (LSPRs).\cite{Pitarke:2006hta}
The potential applications of these phenomena cover many interdisciplinary areas, including nanophotonics,\cite{giannini11,tame13} electronics,\cite{Walters:2010cj,du16}  telecommunications,\cite{kim13,Lu:15} nanomechanics,\cite{soavi16,wang16} photochemistry,\cite{gao11,nordlander14} energy conversion \cite{Li:2015hn,atwater10}
and quantum technology,\cite{gramotnev10,pelton15} since they can be exploited as quantum emitters, nanoantennas, optical detectors, sensors, data storage, and energy harvesting devices.
It is thus clear why the research of plasmonic materials  active in the IR-visible range has fostered such a huge effort in the fields of physics and material science.

An ideal plasmonic material should have tunable plasma frequency, low energy loss, high chemical, mechanical and thermal stability, low cost and high integrability with existing  CMOS (complementary metal-oxide semiconductor) technology.
Obviously, gathering together all these properties in a single material is not trivial, and depending on the specific application specific features are preferred. 
For example, because of their plasmonic response
in the visible range, the high chemical stability, and biocompatibility, noble metals (especially Au and Ag) are considered the plasmonic materials {\em per excellence}, and have been largely used 
for biological sensing, \cite{C3CS60479A} waveguiding,\cite{Oulton08} energy transfer processes,\cite{Cushing12} light-harvesting.\cite{Linic11}
However, the high negative permittivity and high energy losses prevent their use as hyperbolic metamaterials in the visible range; moreover  
the chemical softness and the low melting points make noble metals unsuitable for applications in high temperature and strong light illumination regimes,
such as thermophotovoltaic\cite{fan14} and photothermoelectric\cite{Kraemer11,Catellani:2015by}, heat-assisted magnetic recording\cite{Gage09} systems.
Thus, while the tunability of the plasma frequency of noble metals can be partially improved by changing the geometry, the size and the intermetallic composition of the samples, other properties related to the thermal and mechanical stability or to the growth conditions impose to find alternative materials.\cite{Naik:2013hh}

In order to overcome these problems, a new class of refractory ceramics (e.g. TiN, TiC, ZnN) has been proposed as promising plasmonic materials in the visible and infrared (IR) range.\cite{boltasseva14} 
Being refractory these compounds exhibit an extraordinary mechanical stability over a large range of temperatures ($\sim2000^{\circ}$C) and pressures ($\sim$3.5 Mbar), well above the melting point of standard noble metals ($\sim800^{\circ}$C). These materials are also resistant to corrosion and compatible to silicon technology. 
Among these, titanium nitride has optical and plasmonic properties (color, electron density, plasmon frequency) very similar to gold and has been exploited  for the realization 
of waveguides,\cite{Naik:2012te}  broadband absorbers,\cite{Li:2014bj} local heaters,\cite{Guler:2013iw} and 
hyperbolic metamaterials,\cite{Naik:2014hl,Saha:2014gx} in connection with selected dielectric media (e.g. MgO, AlN, sapphire, etc).

Even though the fundamental mechanical and optoelectronic properties of TiN have been largely studied so far from experimental and theoretical point of view,\cite{pearson93,exp2}
very little is know about its plasmonic behaviour. 
Here, we present a fully first-principles investigation, based on time dependent density functional theory (TDDFT), of the plasmon properties of stoichiometric titanium nitride.
The microscopic origin of plasmonic excitations in the different range of the electromagnetic spectrum are analyzed in terms of the fundamental collective and/or radiative excitation of  TiN electronic structure. 
In particular, from the simulation of energy loss spectra at different momentum transfer we 
derivate the TiN plasmon dispersion relations  that are a fingerprint of the material and are directly accessible by experimental measurements. 
Further, the comparison between simulated dispersion curves and ideal free-electron model allows us to 
investigate the origin of the dissipative behavior of the material, which is an unavoidable prerequisite for any
realistic application.
We further demonstrate the retention of TiN optical properties against the applied pressure and we prove a universal scaling law that relates the bulk modulus and the plasmonic properties of TiN.
Finally, we present the analysis of the surface-plasmon polaritons of TiN when interfaced with dielectric materials as proposed for the realization of hyperbolic metamaterials and waveguides.
The similarities and the differences with other noble metals, in particular with gold, are thoroughly discussed all along the paper.

\section{Computational Details}\label{method}

Electronic structure calculations were carried out by using a first principles total-energy and forces approach based on density functional theory (DFT), as implemented in the Quantum-Espresso suite 
of codes.\cite{Giannozzi:2009p1507} PBE\cite{pbe} generalized gradient approximation was adopted for parameterisation of the exchange-correlation functional. 
Ionic potentials were described by ultrasolt pseudopotential of Vanderbilt type.\cite{uspp}  Ti($3s$) and Ti({$3p$)
electrons were  explicitly included in the valence shell. Single particle wavefunctions (charge) were expanded in plane waves up to a kinetic 
energy cutoff of 30 Ry (300 Ry). A uniform mesh  of $(24\times24\times24)$  k-points was used to sample the 3D Bruillouin zone of TiN bulk. 

Low-index TiN surfaces were simulated through periodically repeated slabs,
each including 21 layers of TiN, with different lateral periodicity and orientation (namely (100), (110), Ti- and N- terminated (111)),
separated in the direction perpendicular to the surface by a 12~\AA ~ layer of vacuum, to avoid spurious interactions among replica. 
Preliminary tests with large supercells 
did not evidence any surface reconstruction, thus 
surface calculations were then performed in a $(\sqrt2 \times \sqrt2)$ 2D unit-cell  for (001) and (110) surfaces, and in a $(1 \times 1)$ 2D unit cell for the Ti- and N-(111) surfaces.
Accordingly, uniform 2D meshes of $(9\times9)$, $(9\times16)$, and $(12\times12)$ k-points were used for
(100), (110), and (111) surfaces, respectively. 

EELS spectra were calculated by using the {\sc turboEELS} code,\cite{turbo_eels} included in the Quantum-Espresso distribution. 
{\sc turboEELS} implements a Liouville-Lanczos approach to linearized time-dependent density-functional theory, designed to simulate electron energy loss and inelastic X-ray scattering spectra in periodic systems.\cite{eels}
Within the linear response approximation, this approach allows one to simulate the loss function $L=-Im[\hat{\epsilon}^{-1}(\bf{q},\omega)]$, where $\hat{\epsilon}( {\bf q},\omega)=\epsilon_1+i\epsilon_2$ is the complex dielectric function, 
 {\bf q} is the transferred momentum and $\omega$ is the frequency. 
Crystal local field effects  (CLFEs) and exchange-correlation local field effects (XCLFs) \cite{PhysRevB.81.085104} are both explicitly taken into account in the {\sc turboEELS} code. CLFEs are particularly relevant for large transmitted {\bf q} and account for  non-homogeneity of the system on the microscopic scale.
 The optical response function  $\hat{\epsilon}( {\bf 0},\omega)$, is obtained in the limit $ {\bf q}\rightarrow{\bf 0}$. We carefully checked that for values of $| {\bf q}| \le 0.015$\AA$^{-1}$ the TDDFT spectra exactly reproduce the results obtained solving an independent-particle formulation 
of the frequency-dependent Drude-Lorentz model for the dielectric function $\hat{\epsilon}(\omega).$\cite{PRB_TB}

The electronic properties of selected dielectrics (MgO, AlN, GaN, CeO$_2$),  considered later in the text as prototypical TiN/dielectric interfaces,
have been calculated ab initio by using a recent 
pseudohybrid Hubbard implementation of DFT+U\cite{acbn0} that profitably corrects  the energy bandgap\cite{acbn0,Gopal:2015bf} as well as the dielectric and vibrational properties of semiconductors.\cite{Calzolari:2013kv}
A novel first principles tight-binding approach\cite{PRB_TB} was used to  simulate the optical properties of dielectrics
through the  solution of the generalized Drude-Lorentz expression of the macroscopic dielectric function $\hat{\epsilon}(\omega)$
where both intraband (Drude-like) and interband (Lorentz-like) contributions are explicitly taken into account.

The simulation of the color is based on the {\em tristimulus} colorimetry theory,\cite{colorimetry,color} which provides a red (R), green (G), and blue (B) representation of the perceived hue, starting from the knowledge of an illuminant source (tabulated), the retina matching functions (tabulated) and the calculated absorption coefficient  $\alpha(\omega)\propto\omega\epsilon_2(\omega)$.

\section{Results and discussion}\label{results}
\subsection{TiN bulk - optoelectronic properties}

As a preliminary step, we characterized the electronic and optical properties of stoichiometric TiN bulk at zero temperature and zero applied pressure. 
At standard conditions, the differences in electronegativity and atomic size between N and Ti give rise to a 
compound that although metallic still preserves a partial ionicity in the bonds. The commingling of these two effects is responsible for most of the peculiar features of TiN. 
As many ionic compounds, TiN crystallizes in a close-packed rocksalt  structure, where nitrogen occupies octahedral interstitial
sites in the ({\em fcc}) Ti sublattice. The simulated lattice parameter $a_0$=4.244 \AA~ and the bulk modulus $B_0$=2.7 Mbar (Table \ref{tab1}) perfectly reproduce the experimental values $a_0$=4.242 \AA~ and $B_0$=$2.9$ Mbar, respectively. \cite{pearson93,exp2}

\begin{figure}[!h!]
\begin{center}
\includegraphics[width=0.9\columnwidth]{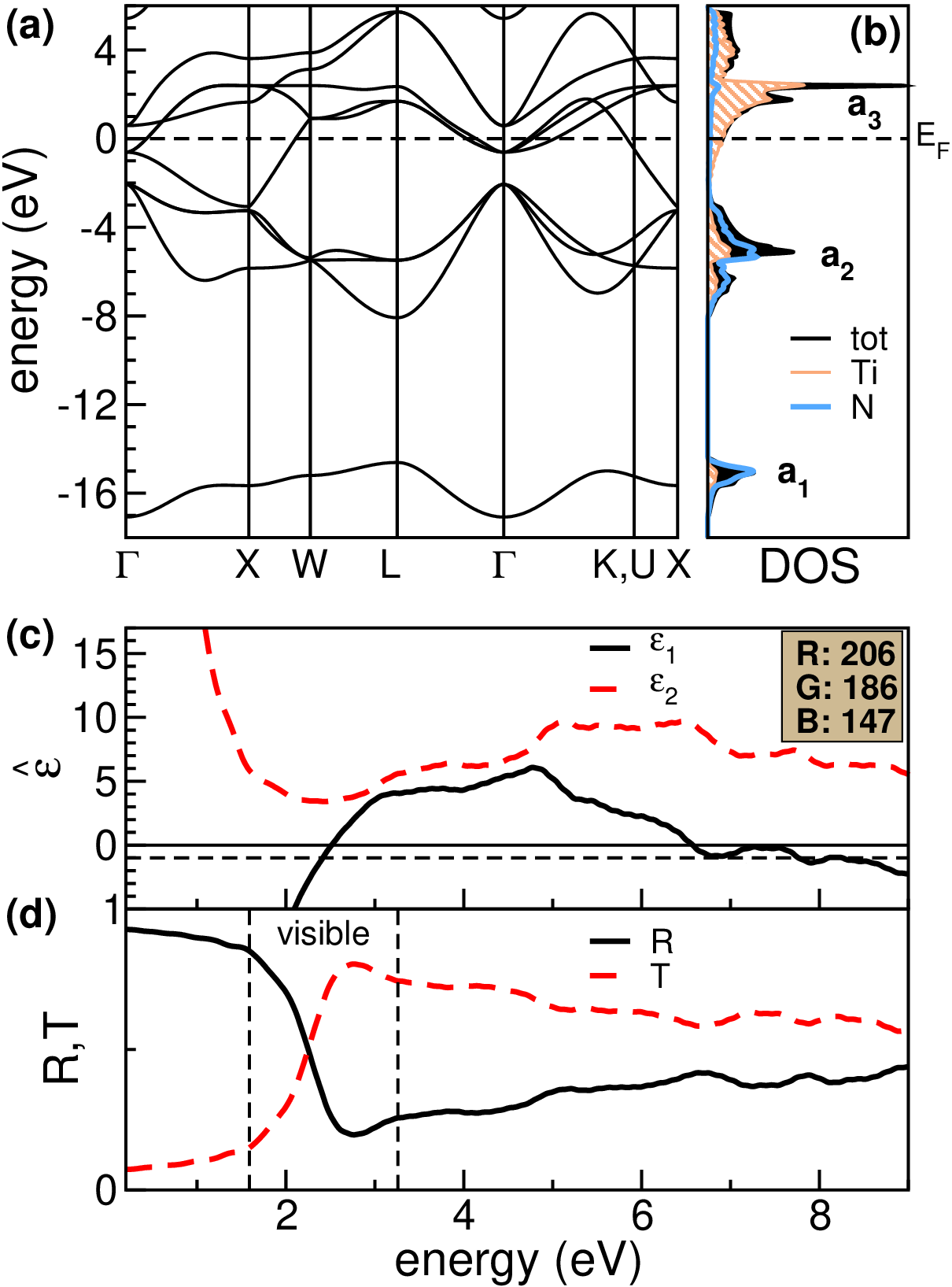}
\vspace{5mm}
\caption{\small  Color online. Electronic and optical properties of TiN bulk at equilibrium (NaCl structure, P=0 Mbar). (a) Bandstructure.  (b) Total (black area),  Ti- (orange dashed line) and N-projected (blue straight line) DOS.  
(c) Real ($\epsilon_1$, black straight line) and imaginary ($\epsilon_2$, red dashed line) components of the dielectric function. Inset shows the simulated color; the corresponding RGB code is also indicated.
(d) Reflectivity (R, black straight line) and transmittivity (T, red dashed line) spectra. 
The zero energy reference in panels (a,b) is set to the Fermi level (E$_F$, dashed line). 
Horizontal straight (dashed) line in panel (c) identifies the values for which $\epsilon_1$ is zero (minus one), respectively.
Vertical lines in panel (d) indicate the visible range.}
\label{fig1}
\end{center}
\end{figure}

\begin{table}[!b!]
\caption{Structural and plasmonic parameters of TiN bulk at different applied pressure for NaCl-like (B$_1$)
and CsCl-like (B$_2$) crystalline phases. Pressure ($P$) and bulk modulus (B$_0$) are expressed in Mbar, plasmon energies ($E_{p}$) are given in eV.
}
\begin{tabular}{c|c|c|c|c}
\hline\hline
phase   & P(Mbar)  & a$_0$ (\AA) & B$_0$ (Mbar) & $E_{p}$ (eV) \\  
\hline \hline
B$_1$   &  0.00      &  4.244         &   2.71               &  2.50              \\
             &  0.25      &  4.123         &   3.52               &  2.75              \\
             &  0.50      &  4.045         &   4.38               &  2.88              \\
             &  0.75      &  3.981         &   5.21               &  3.00              \\
             &  1.00      &  3.928         &   6.04               &  3.09              \\
             &  1.25      &  3.874         &   6.99               &  3.15              \\
             &  1.50      &  3.837         &   7.61               &  3.24              \\
             &  2.00      &  3.765         &   9.11               &  3.34              \\
\hline
B$_2$  & 3.50        & 2.229         &    -                   & 1.90             \\   
\hline\hline                  
\end{tabular}\label{tab1}
\end{table}

The electronic and optical properties of TiN are summarized in Figure \ref{fig1}.
The mixed covalent-ionic character of TiN reflects in the bandstructure (panel a) where we identify 
3 separate sets of bands, each with a predominant character deriving from a single constituent (ionic contribution) plus a minor contribution 
from the other element (covalent contribution), as evident from the total (black area)
and projected (straight and dashed lines) density of states (DOS, panel b).
At higher binding energies [-17.0,-14.5] eV  we found a single band, with a primary N($2s$) character, marked as $\bf{a_1}$ in Figure \ref{fig1}b.
The energy range from  -8.0 to -2.0 eV is characterized by a set of three dispersive bands (labeled  $\bf{a_2}$), degenerate at $\Gamma$, which
mostly derive from N($2p$) states only partially hybridized with Ti($e_g$) orbitals. Finally,  a bundle  of five overlapping bands (labeled $\bf{a_3}$) spans
the energy range [-2.0, 4.0] eV, around the Fermi level ($E_F$, set at zero energy). 
These bands are responsible for the metallicity of TiN and have a net Ti($t_{2g}$) character with a weak  contribution from N($2p$) states.
The three lowest-energy bands of the bundle, degenerate at $\Gamma$, cross the Fermi level at several points of the Brillouin Zone (BZ), giving rise to a multi-pocket Fermi surface.
The sharp peak in the DOS (panel b) at $\sim$2 eV above the Fermi level originates in one of these three bands that  run almost flat for a large portion of the BZ (e.g X-W-L directions). Our results reproduce previous bandstructure calculations obtained with different  techniques, such as tight-binding models \cite{Margine:2011kr} or first principles full-potential linear muffin-tin orbital \cite{delin96,ahuja96} and full-potential linearized augmented plane wave \cite{,Stampfl:2001gn} density functional methods. 
Further, the present results perfectly agree with the experimental features from 
UV and X-ray photoemission and Auger spectroscopies,\cite{Hochst:1982vs, walker97} which identify a nitrogen $2s$ band at $\sim$16 eV binding energy, a double-peaked N($2p$)-Ti($3d$) contribution at [7.0, 3.0] eV binding energy and a metallic band arising at $\sim$2.5 eV binding energy up Fermi level.

The optical properties of TiN are characterized by the complex dielectric function at {\bf q}={\bf 0}, i.e. $\hat{\epsilon}(\omega)=\hat{\epsilon}({\bf 0},\omega)$. 
The calculated real ($\epsilon_1$, straight line) and  immaginary ($\epsilon_2$, dashed line) part of the dielectric function
are shown in Figure \ref{fig1}c.  The imaginary part diverges for $\omega\rightarrow0^+$, this corresponds to a finite {\em dc} conductivity of metals ($\sigma=\omega\epsilon_2/4\pi$), in agreement with experimental results.\cite{Patsalas:2001iw} 
$\epsilon_1$ is generally negative up to $E_{up}$=25.4 eV, which corresponds to the unscreened plasma ($up$) frequency of the system, and reflects the metallic nature of TiN. 
TiN, however,  is not a simple free-electron  metal like Na or Al, since the presence of inter-band transitions  makes $\epsilon_1$ slightly positive in the ranges [2.5, 6.5] eV and [12.0, 18.0] eV (see discussion below).
These features of the dielectric function have been observed in previous theoretical \cite{Kim:2011ib} and experimental works, 
based on different techniques such as optical spectroscopic ellipsometry (SE),
\cite{WIEMER1994181,Patsalas:2001iw,Langereis} optical reflectance\cite{PERRY1988255} 
and  electron energy loss spectroscopy (EELS) with vanishing momentum transfer.\cite{walker97,Pfluger:1984jn}
In particular, Patsalas and coworkers \cite{Patsalas:2001iw}, on the basis of  SE measurements, provided a one-to-one connection between interband contributions to $\epsilon_2$
and optical transitions at high symmetry points of the Brillouin Zone ($\Gamma$, X, L), responsible for absorption peaks at $\sim$ 2.3, 3.9 and 5.6 eV, in perfect agreement
with the multi-structured positive peak in the $\epsilon_2$, shown in Figure \ref{fig1}c.  
In an alternative way, J. Pfluger et al. \cite{Pfluger:1984jn} converged at the same results, interpreting these inter-band contributions in terms of a semiempirical model, where 
the low-energy tail of a Lorentz oscillator - centred at E=7 eV  and associated to p-d transitions - is responsible to the positive values of $\epsilon_1$ in the range [2.5, 7.0] eV. 

A positive $\epsilon_1$ along with a non-zero $\epsilon_2$ in the visible range implies that the compound can absorb light in the blue-region of
the spectrum, imparting to TiN its characteristic gold-like color,  as shown in the  inset of panel 1c. 
The behaviour of the dielectric function affects also the reflectivity (R) and the transmittivity (T) of the sample. At normal incident  conditions R and T are defined as:
\begin{eqnarray}
R &=& \frac{(1-n)^2+k^2}{(1+n)^2+k^2}\\
T &=& 1-R,\nonumber
\end{eqnarray}
where
\begin{eqnarray}
n &=& \{ \frac{1}{2}[(\epsilon_1^2+\epsilon_2^2)^{1/2}+ \epsilon_1] \} ^{1/2}\\
k &=& \{ \frac{1}{2}[(\epsilon_1^2+\epsilon_2^2)^{1/2}- \epsilon_1] \} ^{1/2}\nonumber
\end{eqnarray}
are the real (n) and imaginary (k) part of the complex refractive index $\hat{N}=n+ik$.
From Figure \ref{fig1}d, R is almost 1 in the IR and in part of the visible region, then it  decays abruptly in correspondence of the energy $E_{p}$=2.5 eV, for which $\epsilon_1$ becomes positive. Thus, for  $E>E_{p}$  light can pass through the material, which is no more perfectly reflecting. 
Experimental energy loss experiments \cite{Pfluger:1984jn} confirm the high TiN reflectivity up to 2.5 eV.
The fact that T does not tend to 1 but rather 
decreases at higher energies (e.g. 4-9 eV) indicates that the TiN is not fully transparent in the UV. 
This is a further indication that TiN is not a free-electron metal, but a partially lossy material where absorption processes take place over a large section of the electromagnetic spectrum.

\begin{figure}[t!]
\begin{center}
\includegraphics[width=0.9\columnwidth]{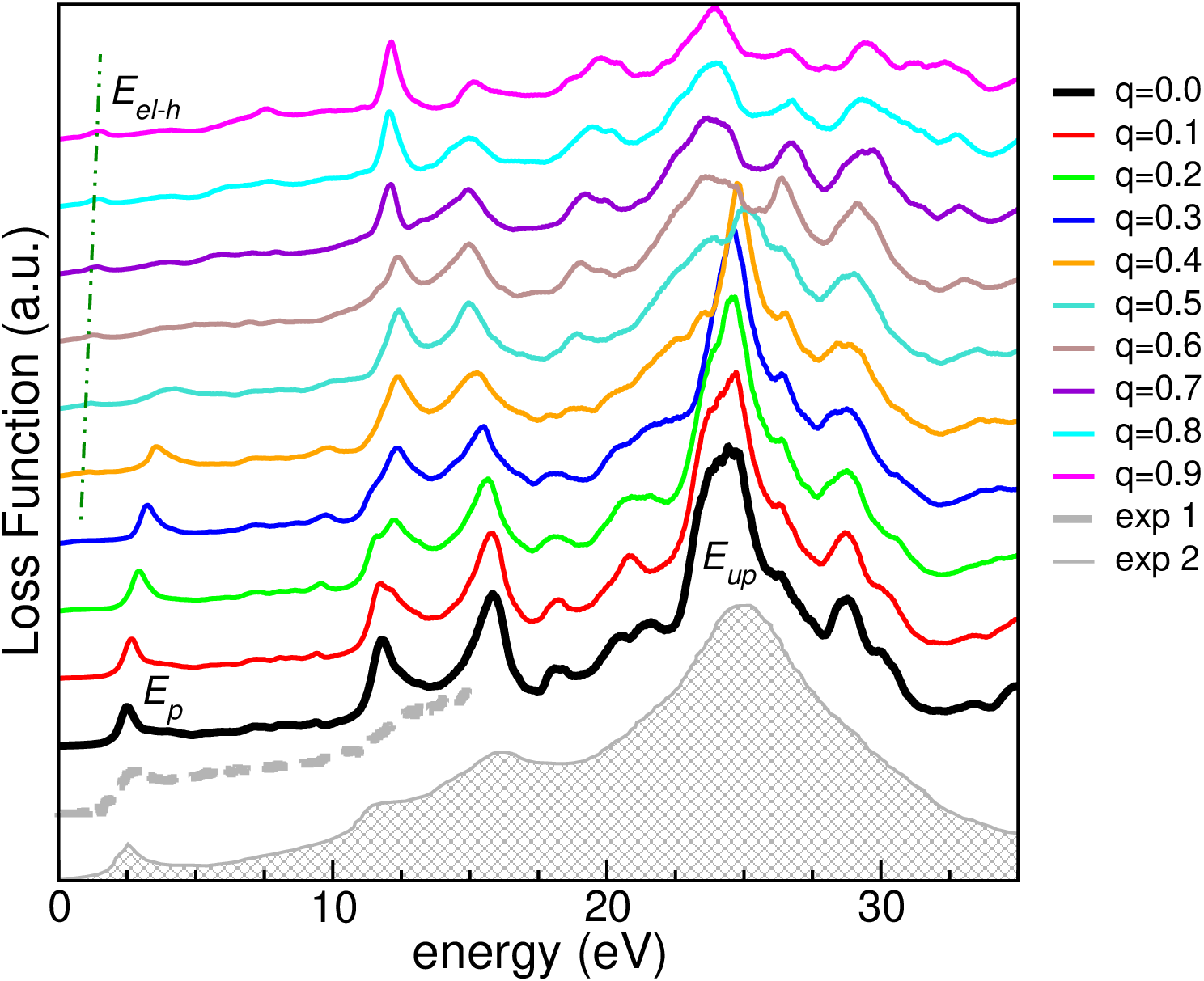}
\vspace{5mm}
\caption{\small  Color online. Simulated loss function for TiN bulk at increased transferred momentum $|${\bf q}$|$ in units of $2\pi/a_0$. Dot-dashed line follows the low-energy single-particle ({\em el-h}) contribution at increasing $|${\bf q}$|$. Two sets of experimental EELS spectra  (gray curves) are superimposed for comparison. Experimental set 1 is adapted from Ref. \cite{Herzing:2016io}, set 2 from Ref. \cite{Pfluger:1984jn}}
\label{fig2}
\end{center}
\end{figure}

\subsection{TiN bulk - plasmonic properties}
The simulated (TDDFT) energy loss functions  for increasing momentum transfer {\bf q} are shown in Figure~\ref{fig2}, along with two independent sets of experimental data for {\bf q}={\bf 0}, collected with EELS-TEM apparatus in ultra high vacuum,\cite{Pfluger:1984jn,Herzing:2016io} superimposed for a direct comparison. 
The dominant feature in  loss function spectra for {\bf q}={\bf 0} (black line= simulation, gray lines=experiments) is a broad peak at $E_{up}$=25.4 eV, which corresponds to the volume plasmon.  
At this energy, the incident electric field is so high that the excited sample electrons behave as free (unscreened) electrons, no more screened by the crystal field.
Thus, we can express E$_{up}$ in terms of the Drude-like  
plasma frequency  $\omega_{up}=E_{up}/\hbar=\sqrt{\frac{e^2n_e}{\epsilon_0m_e}}$,
where $e$ is the electron charge, 
$\epsilon_0$ the dielectric permittivity of vacuum and $m_e$ the electron mass.
By inverting  $\omega_{up}$ 
we can extract the total free-electron density $n_e=4.7 \times 10^{23}$ e/cm$^3$, in agreement with experimental values ($n_e=4.5 \times 10^{23}$ e/cm$^3$).\cite{Pfluger:1984jn}

Apart from this main peak, we identify other plasmonic features at lower energies. Particularly relevant for optoelectronic applications is the plasmon excitation in the
visible range that we found at $E_{p}$=2.5 eV,   very close to the experimental one at
2.8 eV.\cite{Pfluger:1984jn,Herzing:2016io} The origin of this resonance derives from the interplay between inter- and intra-band transitions that can be easily
interpreted in terms of the electronic structure of Figure~\ref{fig1}. For energies lower than 2.5-3.0 eV, the only possible excitations of the valence electrons are due to intra-band transitions through the Ti($3d$) bands
that cross the Fermi level (bands ${\bf a_3}$). A pure $d\rightarrow d$ optical transition would be symmetry forbidden (i.e. vanishing $\epsilon_2$); 
in this case, the partial coupling with the N($2p$) states slightly relaxes the $\Delta\ell$ sum rule, but still gives very low contribution to optical absorption, i.e. not zero but still small $\epsilon_2$. 
As the excitation  energy is increased, inter-band transitions from N($2p$) to Ti($3d$) take place (${\bf a_2}\rightarrow{\bf a_3}$). 
This creates a dielectric screening, which results in the zero cross-over from negative to positive of $\epsilon_1$
at $E_{p}$=2.5 eV. At this energy $\epsilon_1=0$ and $\epsilon_2$ have a minimum (see Figure \ref{fig1}c), thus as a consequnce the loss function has a peak.
This is interpreted as a {\em screened plasmon}, which involves the collective oscillation of a reduced charge density $n'_e< n_e$, since only a fraction of the total valence electrons  can be considered as free, the rest being effectively screened. 

At higher energy the inter-band transitions prevail, the imaginary part (i.e. the absorption coefficient) increases and the loss function rapidly decreases, in agreement with the trend of R and T  in Figure \ref{fig1}d. 
Similar behavior of the dielectric and loss functions characterize the optical  properties of other plasmonic materials, such as noble metals and gold in particular, 
as confirmed by the gold-like color of TiN. However, in the case of noble metals the intraband transitions are due to the $s$-band that crosses the Fermi level, and the inter-band transitions are due $d \rightarrow s$ excitations. In the present case the open $3d$ shell of Ti and the coupling with N further complicates the scenario. In a similar way, the double peaked EELS feature in the [12.0-18.0] eV range, derives
from multiple transitions from ${\bf a_1}$ and ${\bf a_2}$ valence bands to ${\bf a_3}$ conduction bands.

Figure \ref{fig2}  furthermore shows the evolution of the loss function for different  transferred momenta.
All the plasmon resonances undergo a significant broadening and a positive dispersion as the modulus  $q = |${\bf q}$|$ is increased. 
As plasmons are longitudinal excitations of the electronic system, their eigenvalues $E_{p}$ can be derived from the zeros of the longitudinal 
dielectric function $\hat{\epsilon}_L({\bf q},\omega)$. The dispersion relation $E_{p}(q,\omega)$ can be expressed as an expansion of $q$ as:
\begin{equation}
\label{disp}
E_{p}(q,\omega)=E_{p}^0+\alpha q^2 + \beta q^4 + \mathcal{O}(q^4),
\end{equation}
where $E_{p}^0$ is the plasmon energy for q=0. 
Depending on the model used to describe $\hat{\epsilon}_L({\bf q},\omega)$ (such as Lindhard, jellium model, RPA, etc)\cite{Egerton:2011bj} the range of validity of Eq. \ref{disp} and the physical meaning/expression of coefficients $\alpha$ and $\beta$ may be slightly different, since they include different amount of the kinetic and exchange-correlation contributions. \cite{serra91} 
For instance, in the free-electron approximation the
dispersion relation is  quadratic in the transferred momentum and 
the coefficient $\alpha$  is proportional to electronic polarizability.
Our TDDFT results also follow the theoretical trend of Eq. \ref{disp}: from the fit of the  lowest-energy plasmon resonances $E_{p}$ of Figure \ref{fig2}  
we obtain $\alpha$=4.39 eV\AA$^2$ and $\beta$=-3.25 eV\AA$^4$.
However, even though the parabolic term is dominant (see dashed line in Figure \ref{fig3}), no good fit can be obtained for TiN without the inclusion of the fourth-order term.
Moreover, in the case of TiN the quadratic term $\alpha$ cannot be simply associated to the electron gas polarizability, as it includes 
a non-trivial screening normalization, which derives from the interband transitions.

\begin{figure}[!t!]
\begin{center}
\includegraphics[width=0.9\columnwidth]{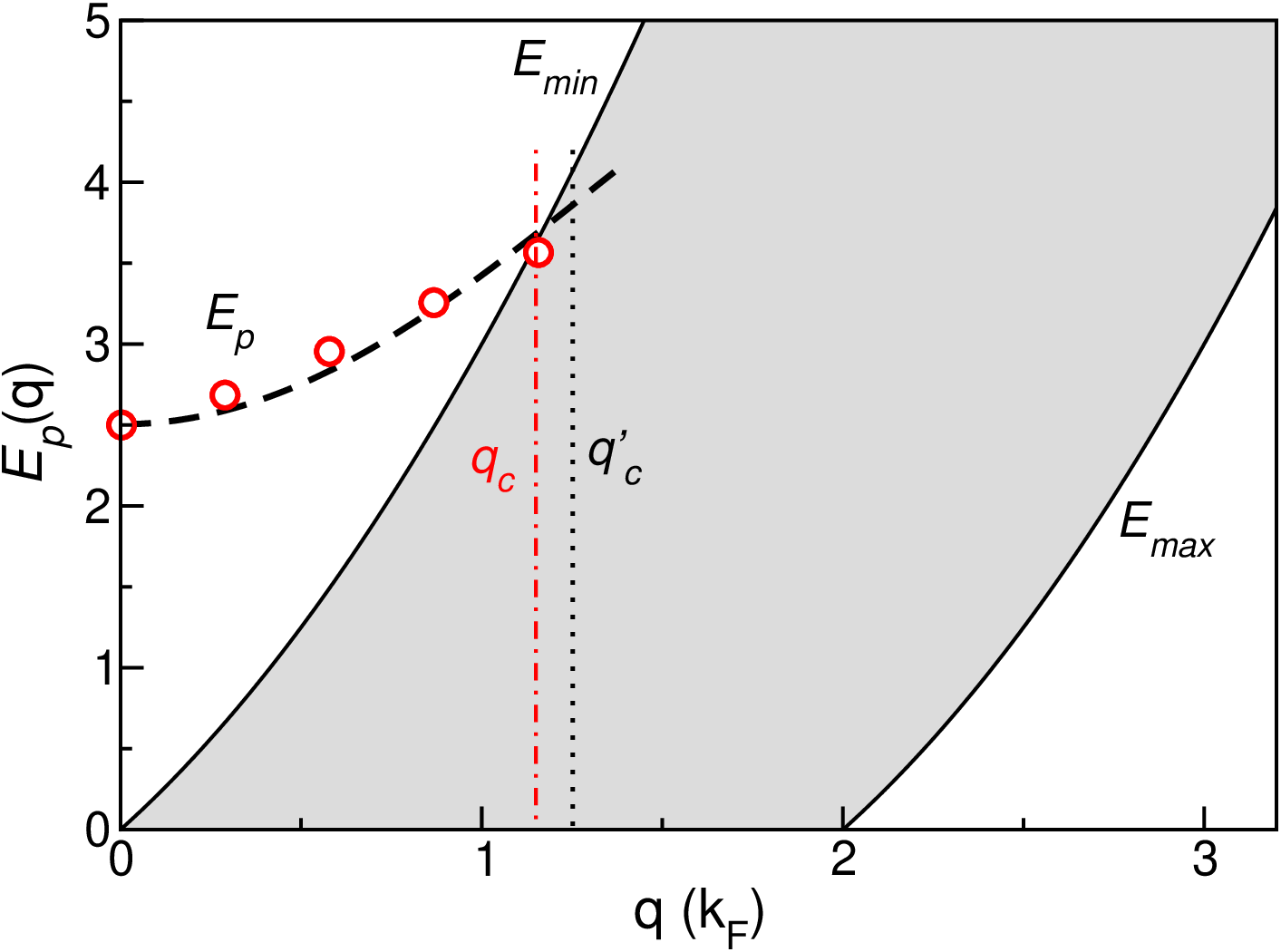}
\vspace{5mm}
\caption{\small Color online. Plasmon energy $E_{p}(q)$ dispersion relations. Red circles are the calculated lowest-energy plasmon resonances $E_{p}$ of Figure \ref{fig2}, and 
black dashed line is the corresponding fit from Eq. \ref{disp}.
$\epsilon_F$ and k$_F$ are the Fermi energy and the Fermi momentum in the free electron model. $q'_c$ (black) and 
$q_c$ (red) are the theoretical  and simulated critical momenta, respectively. Gray shaded area covers the range of allowed single-particle excitations.}
\label{fig3}
\end{center}
\end{figure}

The upward energy dispersion goes hand in hand with  the broadening  and the reduction of the intensity of the plasmonic peak.
Figure \ref{fig2} clearly shows that for values of momentum transfer larger than $q_c\approx 0.4 \times 2\pi/a_0$ ($\approx$0.59 \AA$^{-1}$, orange curve), the plasmon peak is quenched, while other features, such as the one marked as $E_{el-h}$ in Figure \ref{fig2}, arise in
different regions of the spectrum. Indeed, low-intensity $E_{el-h}$ contributions  are mostly
due to single-particle interband transitions among ${\bf a_3}$ bands, 
that are optically active only when mediated by an external momentum, while are dark for q=0.

More generally, above a certain {\em critical momentum} $q_c$ plasma oscillations are heavily damped because plasmons can transfer their energy to single
electrons that dissipate energy undergoing electron transitions among the manyfold of bands crossing the Fermi level. In a free  electron gas  model the critical momentum is defined as  $q_c=\omega_{p}/v_F$, where $\omega_{p}=E_{p}/\hbar$ and $v_F$ is the Fermi velocity of the system. 
For a real metal with a complex Fermi surface (such as TiN), $v_F$ does not have a trivial analytical expression. This prevents to predict a theoretical value of $q_c$ to be compared with the one extracted from Figure \ref{fig2}.

In order to overcome this point, we model TiN as a Drude-like system having the reduced screened charge density $n'_e$ which is associated to the calculated $E_p$, instead of the total charge density $n_e$, responsible of the plasmonic peak at $E_{up}$=25.4 eV. 
We remark that this is a crude approximation not justified {\em a priori}, since TiN is not actually an ideal free-electron system. 
However, this assumption allows us to interpret - in terms of simple parameters - the peculiar properties of the plasmon $E_p$, which is the most interesting for plasmonic applications;
albeit the same analysis can not be extended to the overall 
properties of the material. 
From this model, the simulated plasmon relation (red circles in Figure \ref{fig3}) can be discussed in relation to the ideal single-particle region, whose extrema  
$E_{min}(q)=2q/k_F+(q/k_F)^2$ and  $E_{max}(q)=-2q/k_F+(q/k_F)^2$, 
can be derived from the  momentum conservation rules, as described in many textbooks.\cite{wooten}
Indeed, the plasmon decays rapidly into the
electron-hole pair region, while it is stable outside;  the intersection  between the plasmon dispersion curve and the electron-hole region
defines the  critical momentum. 

To pursue further this free-electron model, the Fermi surface reduces to a sphere of radius $k_F=[3\pi^2 n'_e]^{1/3}$,  the screened charge density is $n'_e=\epsilon_0 m_e \omega_{p}^2e^{-2}=4.5\times 10^{21}$ e/cm$^3$ and the 
theoretical critical momentum results  $q'_c=0.64$\AA$^{-1}$. Gray shaded area in Figure \ref{fig3} covers the range of allowed single-particle excitations. The TDDFT plasmon dispersion $E_{p}(q)$ matches quite well the model trend, even though 
the fact that $q_c < q'_c$ indicates once more the relevance of dissipative effects of the interband transitions.

Summarizing, the present analysis confirms that under standard conditions TiN has optoelectronic properties to many extent similar to gold. In particular the low-energy plasmonic  excitation $E_p$ and the low  $\epsilon_2$ (although not zero) characterize TiN as a relatively low-loss plasmonic material, useful for application in the visible range. The overall plasmonic properties result from the complex balance between intraband and interband transitions  that act as screening term of the global electron gas.

\subsection{TiN bulk under pressure}
 
\begin{figure}[!t!]
\begin{center}
\includegraphics[width=0.9\columnwidth]{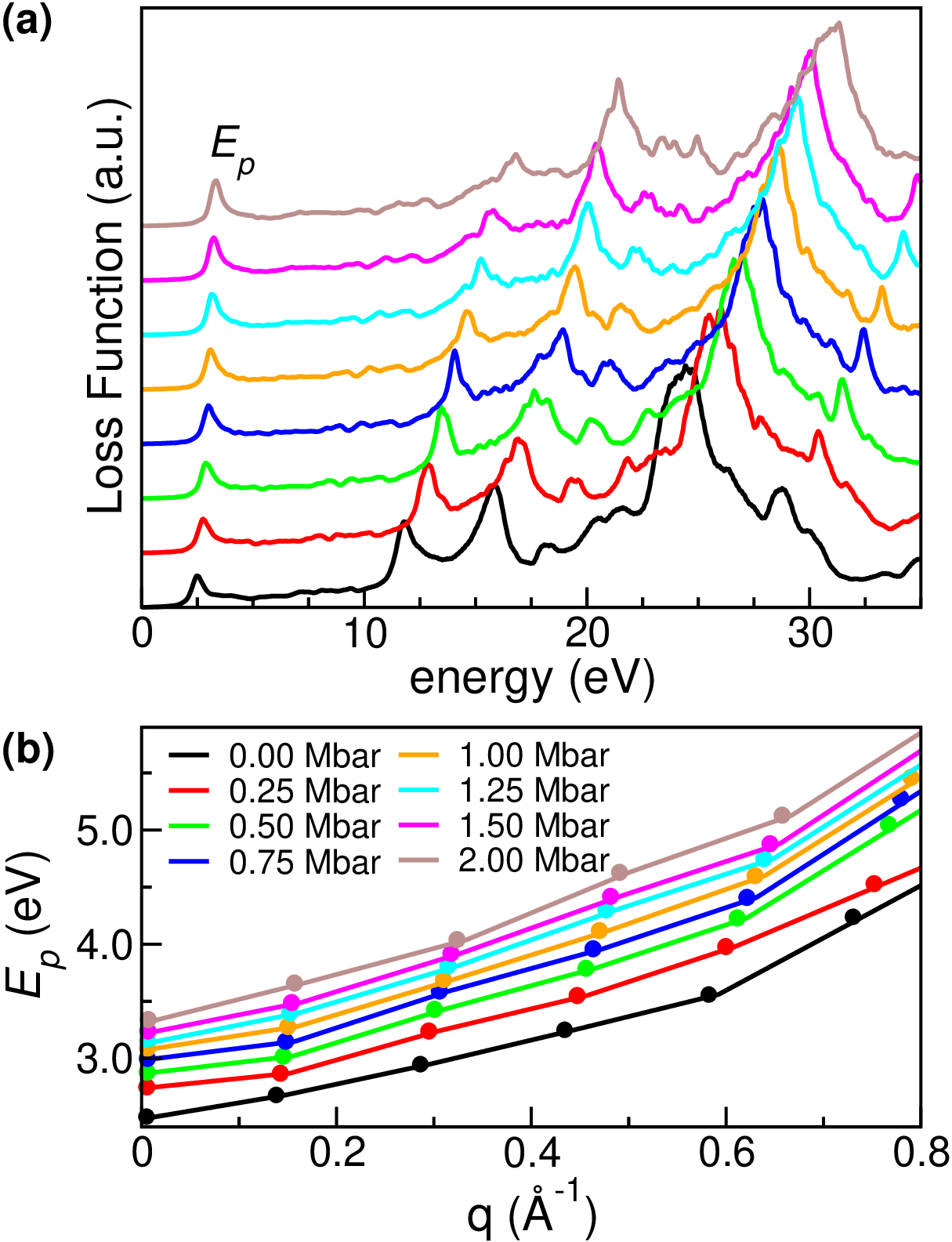}
\includegraphics[width=0.9\columnwidth]{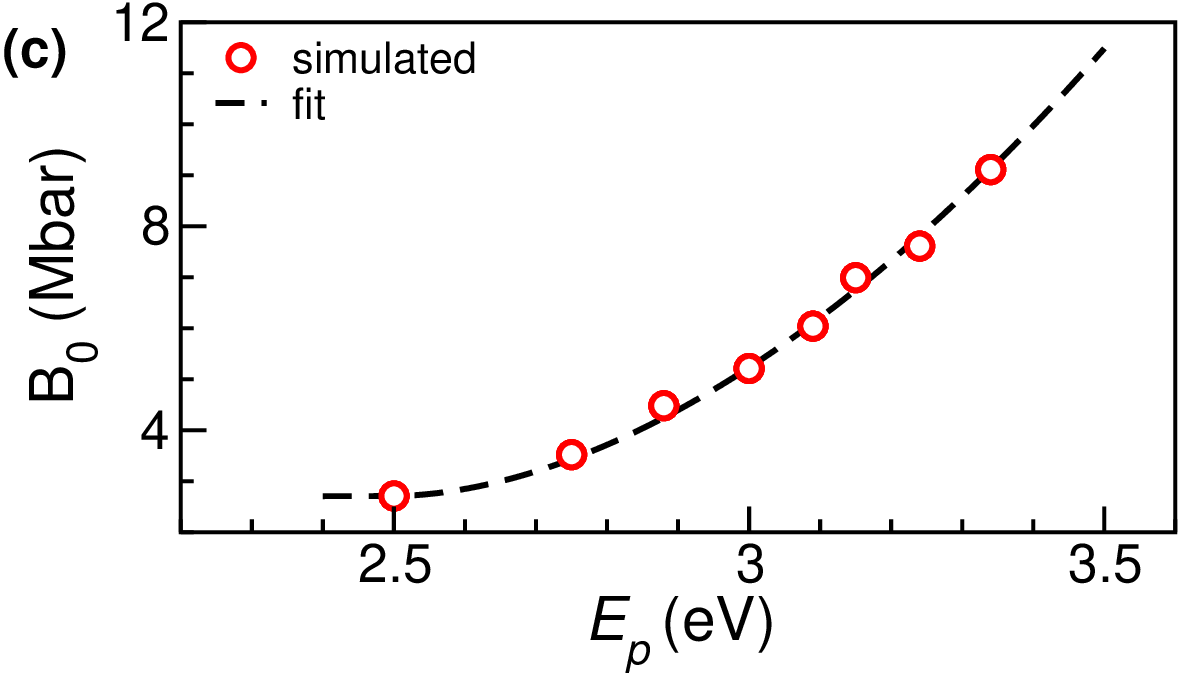}
\vspace{5mm}
\caption{\small Color online. (a)  Loss function spectra for q=0 and (b) plasmon energy $E_{p}(q)$ dispersion for TiN bulk under pressure ($B_1$ phase). Color-pressure code  is  reported in panel (b). 
(c) Bulk modulus B$_0$ plotted against plasmon energy $E_{p}$ at increasing pressure and zero momentum transfer.}
\label{fig4}
\end{center}
\end{figure}

One of the most attracting characteristics of refractory materials is their mechanical resistance along with the retention of the optical properties over a large range
of  applied pressures. 
From the experimental data\cite{pearson93,exp2} we know that 
for pressures $\leq3.5$ Mbar
TiN crystallizes in a rocksalt structure ($B_1$ phase), while at higher pressures TiN undergoes a phase transition to a CsCl-type structure ($B_2$ phase). Phase transition is associated to drastic internal pressure reduction and a consequent volume collapse. 
Here, we do not investigate the phase transition process, which has been largely studied by many authors.\cite{ahuja96,Mishra:2015ev,Yang:2016km} We focus, instead, on the modifications of the plasmon properties as a function of the applied pressure for $B_1$ and $B_2$ structure, separately. 

\begin{figure}[!b!]
\begin{center}
\includegraphics[width=0.9\columnwidth]{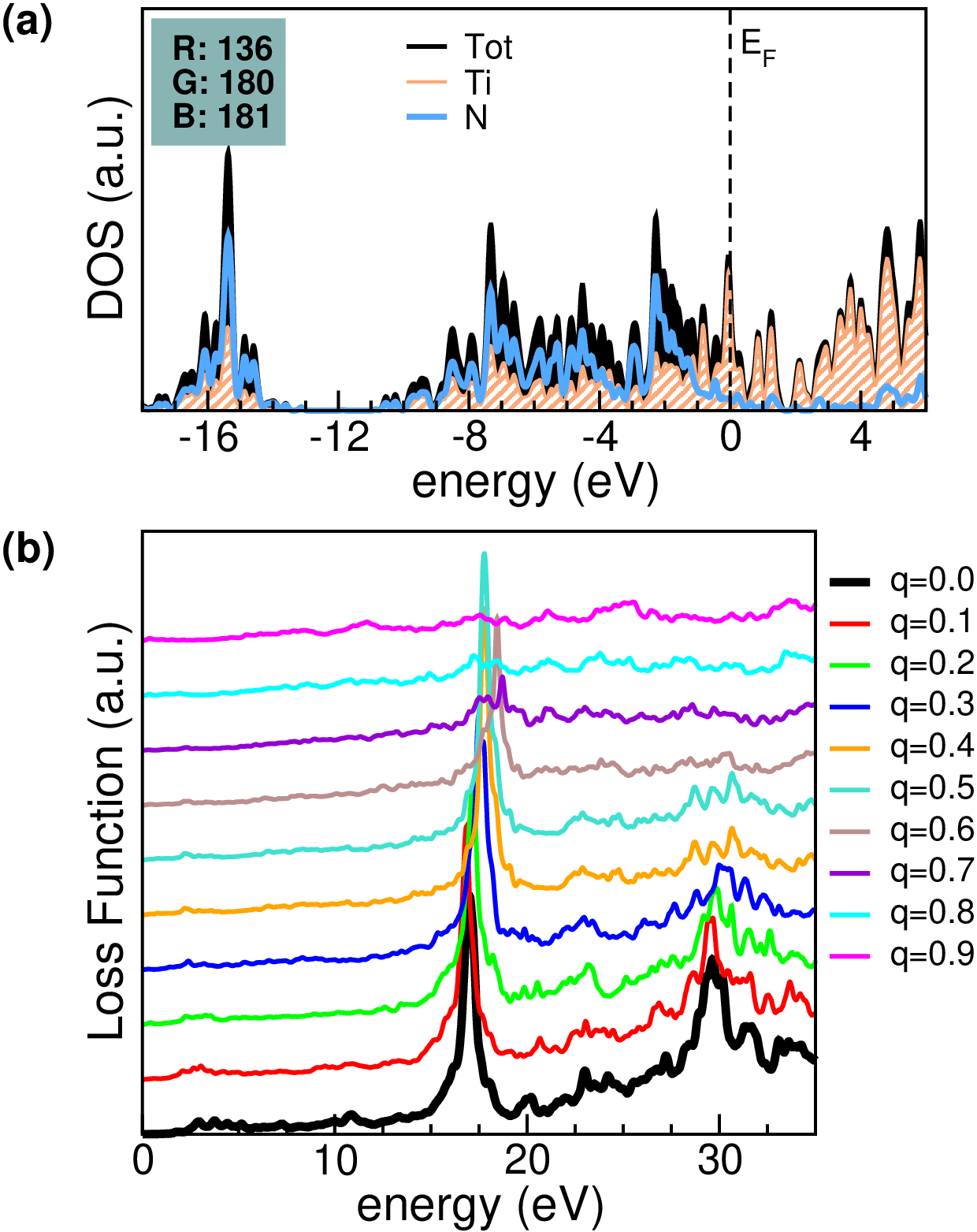}
\vspace{5mm}
\caption{\small  Color online. TiN bulk in B$_2$ phase. (a) Total (black area),  Ti- (shaded orange area) and N-projected (straight line) DOS.
Inset shows the simulated color, the corresponding RGB code is also indicated.
The zero energy reference  is set to the Fermi level (E$_F$, dashed line). 
(b) Loss function  as a function at increased transferred momentum $|${\bf q}$|$ in units of $2\pi/a_0$.}
\label{fig5}
\end{center}
\end{figure}

We first characterized the TiN plasmon dispersion for a set of increasing pressures below the
transition one (namely 0.25, 0.50, 0.75, 1.00, 1.25, 1.50 and 2.00 Mbar), i.e. within  the $B_1$ phase. 
The corresponding variation of structural parameters  (a$_0$ and B$_0$) 
was then obtained
from a set of self-consistent DFT calculations at different volumes. Results are summarized in Table \ref{tab1}.
The increase of $P$ causes the progressive shrinking of the valence band width, while it does not affect the shape and the symmetry of the 
$\bf{a_1}$, $\bf{a_2}$ and $\bf{a_3}$ bands, discussed in Figure \ref{fig1}. The modifications of the electronic structure affect the loss function spectra, as shown in Figure \ref{fig4}a. 
For all the simulated values of $P$, the loss spectra exhibit the same features, albeit shifted at 
higher energies. This is particularly evident for the volume plasmon in the UV, 
as the reduction of the crystal volume with pressure induces
an enhancement of the total charge density $n_e$ and thus of the unscreened peak $E_{up}$. The  blue energy shift is, instead, less
pronounced for the lowest energy plasmon $E_{p}$, since the shrinking of the $\bf{a_2}$ and $\bf{a_3}$ bands partially compensate the
increase of the charge density. The same trend is detected also for the plasmon energy dispersion $E_{p}(q,\omega)$ at each applied pressure (Figure \ref{fig4}b), which maintains the quasi-parabolic behavior discussed above. This confirms that, except for minor details, 
TiN maintains its characteristic plasmonic properties over a wide range of applied pressures, opening the way for a large set of promising 
applications, for which standard plasmonic materials  can not be used.

In their seminal paper, Oleshko and Howe\cite{oleshko07} demonstrated a universal scaling relationship that  
correlates the elastic properties of solid-state materials with (quasi) the square of the plasmon energy. 
Here we extend this relationship to the case of the bulk modulus at different applied pressures.
The plot of bulk modulus B$_0(P)$ against the plasmon energy $E_{p}(P)$ at zero momentum transfer is shown
in Figure \ref{fig4}c. Red circles are the TDDFT simulated data, as reported in Table \ref{tab1}; the dashed line results from the fit of the same data with the power law expression:
\begin{equation}
B_0(P)=B_0(0)+cE_{p}(P)^{\alpha},
\end{equation}
where $B_0(0)=2.71$~Mbar is the bulk modulus at zero applied pressure.
The best fit is obtained for c=8.78 and $\alpha$=1.85, very close to the ideal value $\alpha$=2.0 and in perfect agreement
with the experimental results by Oleshko and Howe ($\alpha=1.9-2.2$).\cite{oleshko07} This is an important results as it can be used as
an analytical tool to quantify 
the modifications of the mechanical properties
{\em via} optical plasmon measurements, in particular in the case of defected and/or substoichiometric  TiN$_x$ compounds.

Finally, we investigated the plasmon properties of TiN in the $B_2$ crystal structure at high pressure (3.5 Mbar). The results are summarized in Table \ref{tab1} and Figure
\ref{fig5}. The transition from NaCl to CsCl structure implies a reduction of the Ti-N  length from 2.12\AA~to 1.93\AA~and  an enhancement
of the covalent contribution to bond. DOS spectra in panel 5a further exhibit  remarkable modifications both in the shape and in the energy
distribution:  the previous peaks ${\bf a_2}$ and ${\bf a_3}$ now largely overlap to form a unique 
broad band that covers the range -11.0 to +5.0 eV.
In particular, a sharp  peak with a net Ti($t_{2g}$) character dominates the energy region across the Fermi level, while a prevalent
N($3p$) peak arises at $\sim$2.0 eV below the Fermi energy. The close presence of these two peaks has a critical effect on the 
optical properties of the system as it drastically lowers the threshold of the inter-band transitions  that now lies at the edge between the infrared and visible ranges. The system thus changes its color from goldish to cyanish (see inset of panel a). As the  absorption spectrum 
has a maximum in the long-wavelength  part of the visible
spectrum, which corresponds to red, the perceived color is dominated by the reciprocal (less absorbed) light colors, i.e. green and blue, whose superposition is, indeed, cyan-like.
The redshift of the absorption edge affects also the plasmonic properties  
 of B$_2$ phase under pressure. 
As a consequence of electron-hole transitions, the loss function at zero momentum transfer
(black line, panel b) is generally broader with plasmonic features less defined than before.  The lowest-energy plasmon is shifted downward at 1.90 eV and its intensity  is strongly quenched with respect to B$_1$ phase. This peak rapidly decays  and it is almost totally quenched for values of $|${\bf q}$|$ as small as $0.1\times 2\pi/a_0=0.28$ \AA$^{-1}$ (red line in panel b). The reduction of the cell volume increases the total charge density of the system,
which corresponds to a blue shift of the unscreened peak from 25.4 to $\sim$30.0 eV.

We conclude that the modifications induced by the pressure to the optoelectronic structure of the B$_2$ phase, are detrimental for application of TiN as
plasmonic  material in the IR-vis range under as-high pressure conditions.

\subsection{Surface Plasmons and Surface Plasmon Polaritons}

\begin{figure}[!t!]
\begin{center}
\includegraphics[width=0.9\columnwidth]{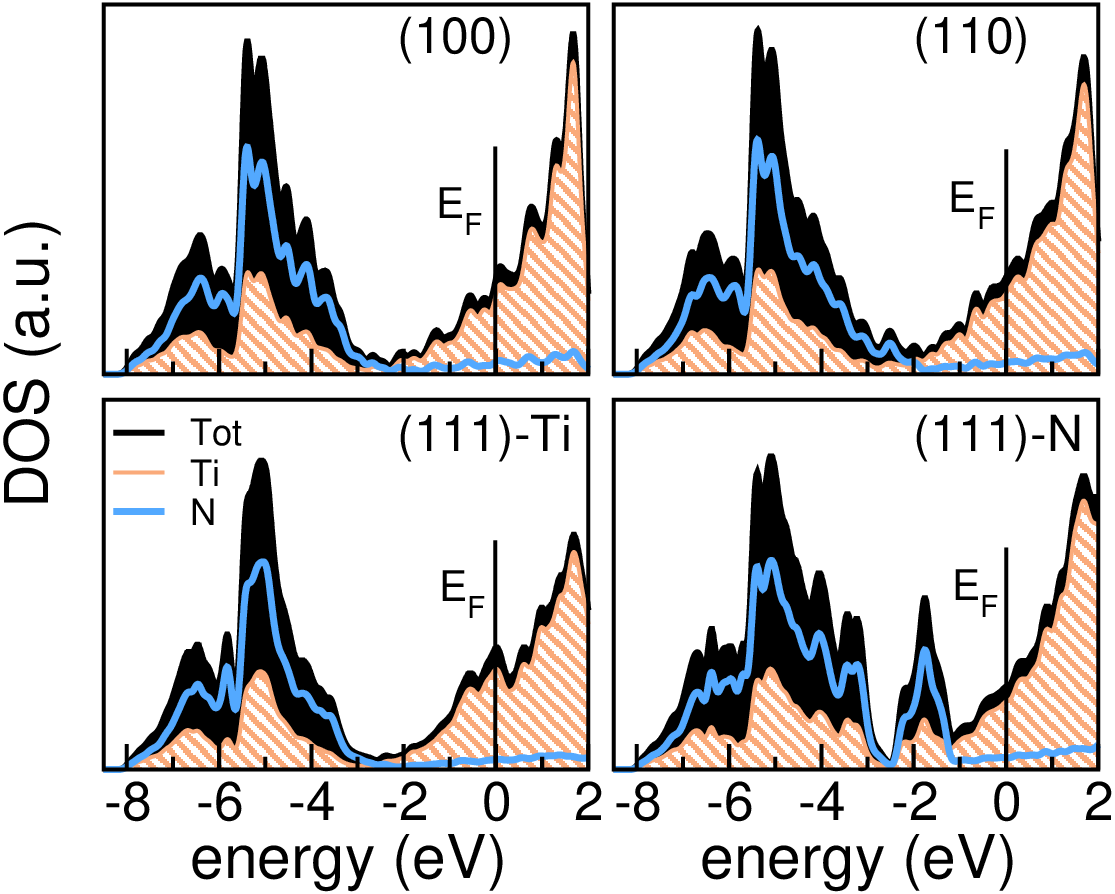}
\vspace{5mm}
\caption{\small  Color online. TiN surfaces. (a) Total (black area),  Ti- (shaded orange area) and N-projected (straight line) DOS. The zero energy reference 
 is set to the Fermi level (E$_F$, black line). }
\label{fig6}
\end{center}
\end{figure}
The existence of surfaces and interfaces modify the optical response of materials. It is thus crucial the study of surface plasmons (SPs), which  are
collective oscillations bound to a metal surface. As they derive from the truncation of the bulk electron gas they are also known as {\em 3D surface plasmons}.
The frequency of SP is a bulk property, while its dispersion is determined from the details of the surface electronic structure.
In general, the energy of SP for a metal/dielectric (m/d) interface can be easily derived from 
classical electromagnetism, by imposing that the electrical and displacement field
must be continuos at the interface.\cite{Pitarke:2006hta} 
This leads to the  {\em nonradiative} SP condition 
$\epsilon_m+\epsilon_d=0$,  
where $\epsilon_{m,d}$ are the  real parts of the dielectric function of the metal and dielectric bulk compounds that form the interface.
In this limit, the surface plasmon does not propagate and does not decay radiatively, but appears as a charge fluctuation at the surface. 
In the case of open surfaces the dielectric is the air,
whose dielectric constant is 1.0, and the condition for SP energy is $\epsilon_m(\omega)=-1$. 

For simple metals, for which the Drude-like expression for the
plasma frequency is valid, the SP energy is reduced to  $E_{sp}=E_{p}/\sqrt{2}$. The energy difference between  $E_{sp}$ and $E_{p}$  usually corresponds to two
well separate peaks in the energy loss spectra. However, for d-metals this simple relation does not hold anymore and the expression $\epsilon_m(\omega)=-1$
must be evaluated numerically. In the case of TiN, if we impose the condition that $\epsilon_1=-1$ (dashed line in Figure \ref{fig1}c) we obtain 
$E_{sp}=2.48$ eV, which fits well the experimental values from reflectivity measurements at low incident angles.\cite{Chen2011}
However, such SP energy is so close to the bulk value (2.5 eV) to be actually indistinguishable in the experimental spectra. 
This is, again, one of the effects of interband transitions that make the slope of $\epsilon_1$ much more steep than the ideal (Drude-like)  lorentzian one. 
A similar behavior is observed in the case of  Ag surfaces, where a unique loss peak is superimposed to a structureless background due to interbad transitions.\cite{Rocca:1995eo} 

In order to elucidate this statement, we explicitly calculated the electronic structure for a set of low-Miller-index surfaces, namely (100), (110), and (111). 
(100) and (110) faces expose an equal number of Ti and N atoms, while (111) surfaces are terminated either with Ti or with N atoms at both slab ends; no mixed Ti/N terminations of simulated slab were considered, in order to avoid spurious electric fields in the vacuum region of the cell. 
After full atomic relaxation, all structures undergo only  minor relaxations of the outermost layers, in agreement with previous theoretical 
calculations.\cite{Marlo:2000uu,Mehmood:2015dy}

The corresponding electronic  structures are collected in Figure \ref{fig6}, where the similarities with  bulk are evident. \cite{eels_surf} The only 
notable differences regard the (111) surfaces: in the Ti-terminated case there is a net enhancement of the Ti($3d$) contribution at the Fermi level, while 
the N-terminated surface is characterized by  N($2p$) peak at $-2.0$ eV, not observed in the other systems. 
In the latter case, the absence of Ti in the outermost layers slightly reduces the spilling  of charge on the surface. 
The analysis of the projected bandstructures (not shown) indicates the presence of surface states in restricted regions (lenses) 
across the edges of the 2D Brillouin zone, while no surface states are present in the proximity of the centre of the  2D Brillouin zone ($\bar{\Gamma}$). 
In addition, we do not identify any parabolic-like surface state at the Fermi level. This seems to exclude the formation of a 2D electron gas confined to the surface plane, \cite{eels_surf}
and thus the possibility to excite {\em 2D surface plasmons} with acoustic linear dispersion.\cite{Rocca:nat, PhysRevLett.105.016801,0295-5075-90-5-57006}
Dedicated surface EELS experiments are requested to confirm this point.
We can conclude that surface truncation does not substantially modify the main electronic properties of the bulk and in particular the energy position N($2p$)
peak with respect to the Fermi level, which is the main responsible for dissipative interband transitions in TiN.

\begin{figure}[!t!]
\begin{center}
\includegraphics[width=0.9\columnwidth]{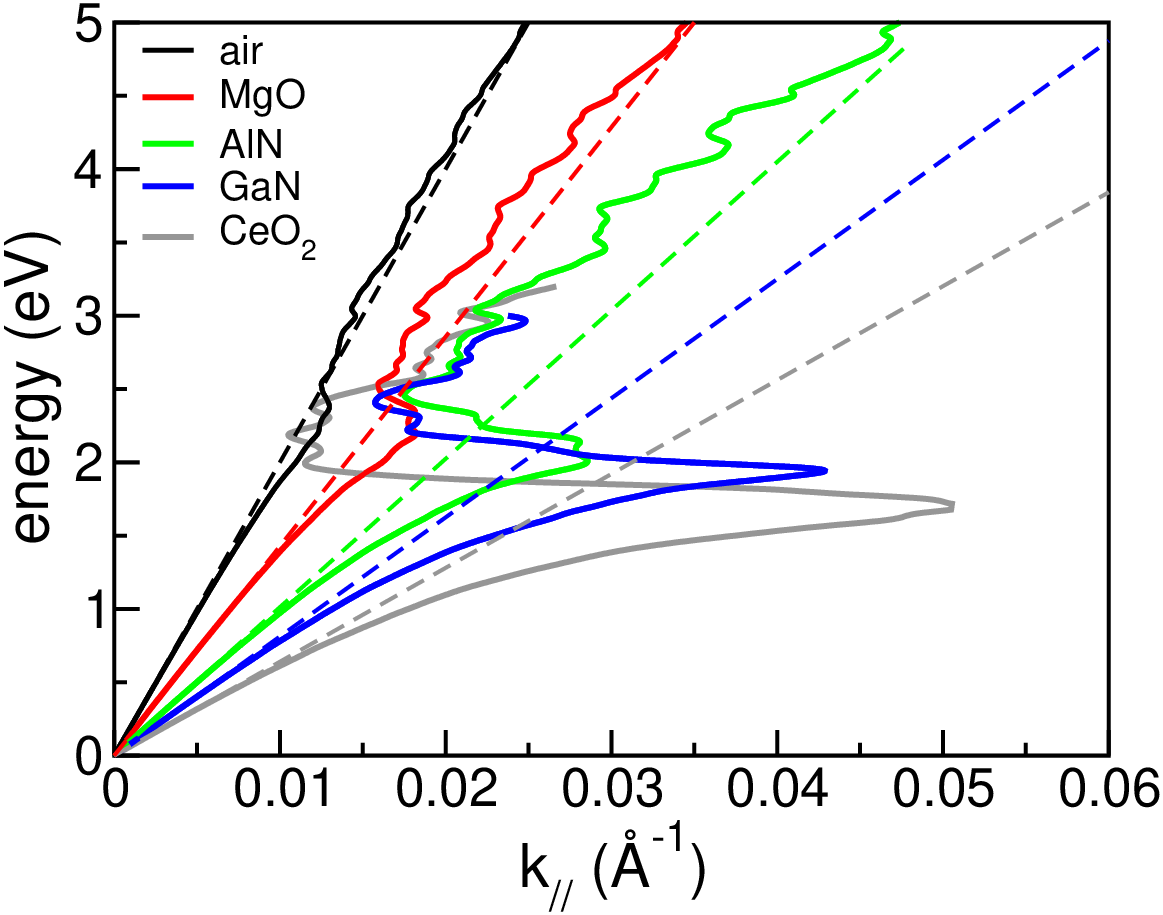}
\vspace{5mm}
\caption{\small  Color online. SPP dispersion relation at TiN/dielectric interfaces. Dashed lines are the dispersion relations for light in the corresponding dielectrics.}
\label{fig7}
\end{center}
\end{figure}

The dissipative contributions to TiN surfaces have also a critical effect on the stability and the spatial properties of 
surface-plasmon polaritons at the TIN/dielectric interfaces. 
SPPs are  mixed oscillatory modes, generated by  coupling 
between  incident light radiation and  surface charge fluctuations.  SSP  can propagate along a metal/dielectric interface
when the condition:
\begin{equation}
k_{\parallel}=\frac{\omega}{c} Re \bigg( \sqrt{\frac{\hat{\epsilon}_m\hat{\epsilon}_d}{\hat{\epsilon}_m+\hat{\epsilon}_d}} \bigg),
\end{equation}
is satisfied,  where $k_{\parallel}$ is the wavevector parallel to metal surface, along the propagation direction,
$\hat{\epsilon}_{m,d}$ are the complex dielectric functions of the metal and the dielectric bulks, and $c$ is the light velocity in vacuum.
For non-dissipative materials $\hat{\epsilon}_{m}$ is real and the SPP dispersion curve $\omega(k_{\parallel})$ has the typical 
behavior with a divergence
at $\omega=\omega_{sp}$ and a forbidden gap for $\omega_{sp}\le \omega \le \omega_{p}$. 
Within this range $k_{\parallel}$ is a purely imaginary number, prohibiting that
the wave propagation exists. For $\omega > \omega_{p}$ (Brewster branch) the metal becomes transparent to the incoming radiation, and the excited wave is no more confined  at the interface. 
For $\omega < \omega_{sp}$ ({\em bound SPP} region) the polaron propagates with a wavevector larger than light in the dielectric at the same energy and the SPP can not decay radiatively emitting photons. At large  wavevectors $k_{\parallel}\gg \omega_{sp}/c$ SPP yields the nonradiative SP limit. 

\begin{table}[!t!]
\caption{Calculated bandgap E$_g$ (eV) and high frequency dielectric constant $\epsilon_d$ of selected dielectric materials. 
}
\begin{tabular}{c|c|c|c|c}
\hline\hline
                      & MgO  & AlN     & GaN   & CeO2       \\  
\hline \hline
E$_g$            & 7.46   &  6.02  &  2.85  &   2.45          \\
\hline
$\epsilon_d$  & 1.8     & 3.8    &  5.9     &  9.5             \\     
\hline \hline                
\end{tabular}\label{tab2}
\end{table}

In the case of TiN the assumption that $\hat{\epsilon}_{m}$ is real does not hold and the dissipative effects described by the imaginary part of the dielectric function must be 
taken into account. Figure \ref{fig7} shows the  dispersion relations of SPP (solid lines) for a set of TiN/dielectric interfaces, along with the dispersion relations of light in the corresponding dielectrics. 
We considered materials that have been eventually
used to realize SPP and metamaterial interfaces with TiN,\cite{Herzing:2016io,Saha:2014gx,Li:2014bj,Naik:2012te,Naik:2014hl,gan} namely 
MgO, AlN, GaN, and we compared the results with the limiting cases of air and CeO$_2$ which have very low and very high dielectric constant, respectively.  The simulated results of  energy bandgap and dielectric constant for these compounds are summarized in Table \ref{tab2}. As the range of interest of TiN/dielectric interface is the Near-IR/visible range,  we assumed  as $\epsilon_d$ the high frequency dielectric 
constant of materials, i.e. without the inclusion of  the LO-TO phonon contribution that is relevant in the far-IR and lowest energy part
of the  electromagnetic spectrum.

The inclusion of the dissipation term strongly changes the character of the dispersion relation.
Although for very low energy, the SPP approaches the corresponding light line, as in the ideal case,
in the vicinity of the surface plasmon energy the dispersion curve bends back toward the light line instead of increasing asymptotically to the surface plasmon energy at infinite momentum. This defines a maximum value ($k_{\parallel}^{max}$) for the excitation of SPP. 
For energy slightly above the dispersion curve of the bound SPP, 
$k_{\parallel}$ is no more purely imaginary and a {\em quasi-bound} SPP exists in place of the ideal plasmon-gap. 
In this energy range, the SPP has a backward group velocity and eventually evolves into the so-called  {\em radiative plasmon} mode  at low $k_{\parallel}$, where it crosses  the photon dispersion line (transition energy). 

The simulated dispersion relations of Figure \ref{fig7} show an increase of the stability of SPP at TiN/dielectric interfaces,
as the dielectric constant is increased.
In the case of air, the SPP dispersion actually does not move away from the light line 
degrading in a radiative mode at very low $k_{\parallel}$, in agreement with the experimental results.\cite{Naik:2012te}
Thus, air is not able to stabilize a SPP at the TiN surface, and the TiN/air interface is not suitable for realistic 
applications.  However, as $\epsilon_d$ is increased, the SPP can  be progressively stabilized, giving rise to 
oscillating waves that propagate along the metal surface, with a wavevector sensitively larger than light (i.e. non radiative mode).
MgO and nitride compounds (AlN and GaN) in particular seem the best choices for the generation of SPP in TiN, in agreement with the
recent experimental studies on these interfaces.\cite{Saha:2014gx,Naik:2014hl,gan} 
Indeed, even though highly dielectric materials such as CeO$_2$
assure a very dispersive energy dispersion relation and large  $k_{\parallel}^{max}$, 
the increase of the dielectric constant sensitively lowers the transition energy, over which the SPP decays radiatively.
The dissipation becomes even more dramatic when the excitation energy overcomes the energy bandgap of the dielectric.
In this case, the dielectric is no more transparent and it absorbs light through valence-to-conduction single particle transitions.
This limits the application potential of  interfaces of TiN with extreme high-k dielectrics, such as ceria.

\section{Conclusions}
We presented a comprehensive theoretical investigation of the plasmonic properties of refractory titanium nitride. Our ab initio results confirm that at standard conditions
TiN exhibits  plasmonic properties in the visible and near-IR regime, very close to gold, in agreement with the experimental data. 
However, in contrast with malleable noble metals, the hardness of refractory
ceramics allows for the exploitation of  plasmonic properties also at high temperature and under pressure.  We have deeply investigated this latter characteristic also in relation to the crystal phase transition, experimentally observed at very high pressure.
The microscopic origin of the plasmon resonances and their dispersions have been discussed on the basis of the analysis of the electronic structure and of the reciprocal interplay between collective and single-particle excitations, which determine the screening and dissipation effects of the electronic system.
We furthermore propose a rule to optimize interfaces with dielectrics for metamaterial applications.

The characteristics of TiN, such as stability, reduced energy loss, and zero-cross over in the IR-visible range, 
promote refractory materials as a promising alternative for the realization
of advanced optoelectronic applications, for which standard plasmonic materials  cannot be used.

\bibliography{biblio.bib}
\end{document}